\documentclass[preprint,12pt]{elsarticle}

\usepackage{graphicx}
\graphicspath{C:\Users\Desktop\QBH-Paper}
\usepackage{epsfig}
\usepackage{amssymb}
\journal{General Relativity and Gravitation}

\begin{document}

\begin{frontmatter}


% %\ead{email address}
\cortext[cor1]{Corresponding Author}

\title{Natural Cutoffs effect on Charged Rotating TeV-Scale Black Hole Thermodynamics}

\ead{msoleima@cern.ch}
\author[a,c]{M. J. Soleimani\corref{cor1}}
\author[b]{N. Abbasvandi}
\author[b]{G. Gopir,}
\author[a,c]{Zainol Abidin Ibrahim}
\author[b]{Shahidan Radiman}
\author[a,c]{W.A.T Wan Abdullah}
\address[a]{National Center for Particle Physics, IPPP, University of Malaya, 50603 KL, Malaysia}
\address[b]{Physics Department, FST, University Kebangsaan Malaysia, 43600 Bangi, Malaysia}
\address[c]{Physics Department, Faculty of Science, University of Malaya, 50603 KL, Malaysia}

\begin{abstract}
We study the thermodynamics of charged rotating black hole in large extra dimensions scenario where quantum gravity effects are taken into account. We consider the effects of minimal length, minimal momentum, and maximal momentum as natural cutoffs on the thermodynamics of charged rotating TeV-scale black holes. In this framework the effect of the angular momentum and charge on the thermodynamics of the black hole are discussed. We focus also on frame dragging and Sagnac effect of the micro black holes.
\end{abstract}


\end{frontmatter}



\section{Introduction}
\label{sec:intro}

The possible of the existence of large extra dimensions \cite{arkani,antoniadis,arkanni2,randall} has opened an exciting field of research on commonly predicts the existence of the quantum gravity. The study of black hole production at particle colliders, such as the Large Hadron Collider (LHC) \cite{emparan,giddings}, and the Muon Collider \cite{www}, as well as in the Ultrahigh Energy Cosmic Ray Air showers \cite{feng,cavagli}, are some of the most significant consequences of TeV-scale quantum gravity. Various approaches to quantum gravity such as quantum geometry \cite{capozziello}, string theory \cite{veneziano,amati,kato}, and loop quantum gravity \cite{garay} all support the idea that near the Planck scale, the standard Heisenberg uncertainty principle should be reformulated by the so-called Generalized Uncertainty Principle (GUP) \cite{kempf,kempf1}. Moreover, some Gedanken experiments in the spirit of black hole physics have also supported the idea of the existence of a minimal measurable length \cite{scardigli}, which is of the order of the Planck length, ${l_p} \sim {10^{ - 35}}m$ \cite{mggiore,hossenfelder}. On the other hand, there is an upper bound for momentum fluctuation based on the context of Doubly Special Relativity (DSR) \cite{ali,das}, and a test particle's momentum cannot be arbitrarily imprecise. This means that there is also a maximal particle momentum, ${P_{\max }}$.
\\In this paper, we are going to study the quantum gravity effects on the thermodynamics of rotating micro black holes with conserved charge during their formation and decay process \cite{calmet} incorporating GUP. In this case, we consider the micro black hole as a semi-classical black hole just with the initial angular momentum of the black hole due to the spin states of the incoming partons. Firstly, we consider a most general form of GUP that admits a minimal length, minimal momentum, and maximal momentum. We study the thermodynamics of charged rotating TeV-Scale black holes with extra dimensions in the Arkani-Hamed, Dimopoulos, and Dvali (ADD) \cite{arkani} scenario in the context of the mentioned GUP. Here, we generalize the thermodynamic parameters of charged rotating micro black holes such as Hawking temperature and Bekenstein-Hawking entropy. We also investigate frame dragging and rotation effects on space-time by the Sagnac interferometrical detection described in terms of the gravito-magnetic Aharonov-Bohm effect. Effects of extra dimensions, mass, angular momentum, and charge of the black hole are also discussed. 
\\The article is organized as follows: Section 2 gives our introduction of a generalized uncertainty principle which admits a minimal length, minimal momentum, and maximal momentum. In section 3, we obtain an expression for charged rotating TeV-Scale black hole Hawking temperature and entropy. Section 4 is devoted to discussing the angular momentum and charge effect on thermodynamics of the black hole. In section 5, we investigate frame dragging effect of the charged rotating micro black hole on space-time by considering the Sagnac effect in the higher dimensional Kerr-Newman fields in the presence of GUP. The paper follows by conclusion in section 6.

\section{Minimal length, minimal momentum, and maximal momentum}

As has been revealed in the introduction, the existence of a minimal observable length of the order of the Planck length is a common feature of all promising quantum gravity proposals \cite{mggiore}. Such a smallest length modifies the Heisenberg uncertainty principle to the so-called Generalized (Gravitational) Uncertainty Principle (GUP) in such a way that $\delta x \ge \frac{\hbar }{2}\left( {\frac{1}{{\delta p}} + {\alpha ^2}l_p^2\frac{{\delta p}}{\hbar }} \right)$, where $\alpha$ is a dimensionless constant of the order of unity which depends on the details of the quantum gravity hypothesis. In the Heisenberg uncertainty principle frame work, $\delta {x_0}$ as the minimal position uncertainty could be made arbitrarily small toward zero due to no essential restriction on the measurement precision of the particle's position. On the other hand, the existence of a minimal measurable length would restrict a test particle's momentum uncertainty to take any arbitrary values leading non-trivially to an upper band, based on the Doubly Special Relativity theories \cite{camelia,amelino,amelino1}. So, there is a maximal particle's momentum due to a fundamental structure of space-time at the Planck regime. Based on the above argument the GUP that predicts such a smallest length and maximal momentum can be written in such a way that
\begin{equation}
\label{2.1}
\delta x\delta p \ge \frac{\hbar }{2}\left[ {1 - 2\alpha {l_p}\left( {\delta p} \right) + 4{\alpha ^2}l_p^2{{\left( {\delta p} \right)}^2}} \right]
\end{equation} 
and this leads us to the following canonical commutation relation,
\begin{equation}
\label{2.2}
\left[ {x,p} \right] = i\hbar \left( {1 - \alpha p + 2{\alpha ^2}{p^2}} \right)
\end{equation}
where $\alpha$ is a positive constant in the presence of both minimal length and maximal momentum. 
\\By generalizing the Heisenberg commutation relation \cite{kempf2}, it was found that for large distances, where the curvature of space-time becomes important, a plane wave on a general curved space-time has no rotation \cite{hinrichsen,zarei}. In fact, there appears to be a limit to the precision with which the corresponding momentum can be described and it can be expressed as a non zero minimal uncertainty in momentum measurement. As a consequence of the small correction to the canonical commutation relation, based on the above arguments of this more general frame work of the GUP, one infers the following expression
\begin{equation}
\label{2.3}
\delta x\delta p \ge \frac{\hbar }{2}\left[ {1 - 2\alpha {l_p}\left( {\delta p} \right) + 4{\alpha ^2}l_p^2{{\left( {\delta p} \right)}^2} + 4{\beta ^2}l_p^2{{\left( {\delta x} \right)}^2}} \right]
\end{equation}
which in the extra dimension scenario based on the ADD model can be written as follows \cite{nozari}
\begin{equation}
\label{2.4}
\delta {x_i}\delta {p_i} \ge \frac{\hbar }{2}\left[ {1 - 2\alpha {l_p}\left( {\delta {p_i}} \right) + 4{\alpha ^2}l_p^2{{\left( {\delta {p_i}} \right)}^2} + 4{\beta ^2}l_p^2{{\left( {\delta {x_i}} \right)}^2}} \right]
\end{equation}
where $\alpha$ and $\beta$ are dimensionless, positive coefficients, which are independent of $\delta x$ and $\delta p$ but generally they may depend on the expectation value of position and momentum. The Planck length in a model of universe with large extra dimensions is defined as ${l_p} = {\left( {\frac{{\hbar {G_d}}}{{{C^3}}}} \right)^{\frac{1}{{d - 2}}}}$, in which $G_d$ is the gravitational constant in d dimensional space-time.  
Based on ADD model, it is given by ${G_d} = {G_4}{L^{d - 4}}$, where L is the size of extra dimensions. 
\\We suppose that operators of position and momentum obey the following commutation relation according to the generalized Heisenberg algebra,
\begin{equation}
\label{2.5}
\left[ {x,p} \right] = i\hbar \left( {1 - 2\alpha p + 4{\alpha ^2}{p^2} + 4{\beta ^2}{x^2}} \right)
\end{equation}
It has been shown that the thermodynamic quantities of a spherical black hole can be obtained by using the standard Heisenberg uncertainty principle \cite{ohanian}. In the same manner, the mentioned natural cutoffs brings several new and interesting implications and modifies the result dramatically by incorporating quantum gravity effects in black hole physics and thermodynamics. In what follows, in order to find thermodynamical properties of charged rotating TeV-scale black hole in the ADD scenario of extra dimensions and its quantum gravitational corrections, we use the mentioned more general form of GUP as our primary input and construct a perturbational calculation.
\section{Thermodynamics of charged rotating micro black hole}
Now, we are going to calculate the thermodynamical properties of charged rotating micro black hole, using the generalized uncertainty principle. In this case, the Kerr-Newman geometry describes the empty space surrounding a charged rotating black hole. Based on the extra dimensions scenario and the ADD model, a natural candidate for the TeV-scale charged rotating black hole is that of a higher dimensional Kerr-Newman solution of the Einstein field equation \cite{ghosh}
\begin{equation}
\label{3.1}
\begin{array}{l}
d{s^2} = \left( {\frac{{\Delta  - {a^2}si{n^2}\theta }}{\sum }} \right)d{t^2} - \frac{\sum }{\Delta }d{r^2} + 2a\\
\times \left[ {1 - \left( {\frac{{\Delta  - {a^2}si{n^2}\theta }}{\sum }} \right)} \right]dtd\varphi  - \sum d{\theta ^2} - \\
\left[ {\sum  + {a^2}si{n^2}\theta \left( {2 - \frac{{\Delta  - {a^2}si{n^2}\theta }}{\sum }} \right)} \right] \times si{n^2}\theta d\varphi {}^2 - {r^2}co{s^2}\theta d\Omega _{d - 4}^2
\end{array}
\end{equation} 
where equation (3.1) is in Boyer-Lindquist coordinates.
Here,
\begin{equation}
\label{3.2}
\begin{array}{l}
\Delta  = {r^2} + {a^2} - \frac{\mu }{{{r^{d - 5}}}} - {Q^2}\\
\sum  = {r^2} + {a^2}co{s^2}\theta 
\end{array}
\end{equation}
the parameters $\mu$ and $a$ are respectively related to the mass (M) and the angular momentum (J) of the black hole via the following relations,
\begin{equation}
\label{3.3}
\begin{array}{l}
M = \frac{{\left( {d - 2} \right)}}{{16\pi }}{A_{d - 2}}\mu \\
J = \frac{1}{{8\pi }}{A_{d - 2}}\mu a
\end{array}
\end{equation}
and,
\begin{equation}
\label{3.4}
\frac{M}{J} = \frac{{\left( {d - 2} \right)}}{{2a}}
\end{equation}
Here, ${A_{d - 2}}$ is the area of a unit (d-2) sphere, which is given by
\begin{equation}
\label{3.5}
\begin{array}{l}
{A_{d - 2}} = \int\limits_0^{2\pi } {d\varphi } \int\limits_0^\pi  {sin\theta co{s^{d - 4}}\theta d\theta } \\
\times \prod\limits_{i = 3}^{d - 4} {\int\limits_0^\pi  {si{n^{\left( {d - 4} \right) - i}}{\theta _i}d{\theta _i} = \frac{{2{\pi ^{\frac{{\left( {d - 4} \right)}}{2}}}}}{{\Gamma \left( {\frac{{d - 1}}{2}} \right)}}} } 
\end{array}
\end{equation}
We would like to draw attention to the fact that $Q$ is related to ${Q'}$ by ${Q^2} = \frac{N}{{N - 2}}{{Q'}^2}$, where ${Q'}$ is a Yang-Mill gauge charge \cite{yang} and for a vanishing $Q = 0$, one recovers the Myers-Perry black hole solution discussed in \cite{myers}. In this case, there are regular inner and outer event horizons, $r_\pm$, which we obtain as a solution of equation (3.2)
\begin{equation}
\label{3.6}
{r^{d - 3}} + \left[ {{a^2} - {Q^2}} \right]{r^{d - 5}} - \mu  = 0
\end{equation}
In fact, similar to the Kerr solutions, in our case the metric has two types of the horizon-like hypersurface: a stationary limit surface and an event horizon. While the surface area of the event horizon has been related to the entropy of a black hole \cite{bekenstein}, that of the stationary limit surface has not been given a physical interpretation. On the other hand, when we are dealing with a charged rotating black hole, the event horizon shrinks, and the inner one appears. Particularly, the thermodynamics associated with the outer event horizon of the black hole is related to the fundamental process of Hawking radiation \cite{wu}. By modeling the black hole as a d-dimensional cube of size equal to its event horizon radius, $r_+$, the position uncertainty, $\delta x$, of the Hawking particle at the emission can be chosen as its Compton wavelength which is proportional to the inverse of the Hawking temperature \cite{medved,zhao,nozari1}. The position uncertainty should then not be greater than a specific scale which is defined as follows \cite{adler,xiang}
\begin{equation}
\label{3.7}
2\zeta  \ge \delta {x_i}
\end{equation}    
where, this imposes constraint on the momentum uncertainty and ${\zeta _d} = \sqrt {r_{{ + _d}}^2 + {a^2}}$ where ${r_{{ + _d}}}$ is the event horizon radius in d-dimensional space-time.
\\From a heuristic argument [40], based on the standard uncertainty principle, the uncertainty in the energy of the emitted particle can be obtained as follows,
\begin{equation}
\label{3.8}
\delta E = c\delta {p_i}
\end{equation}
and by setting $G = c = \hbar = 1$, one deduces the following equation for the Hawking temperature of a black hole based on the large extra dimension scenario
\begin{equation}
\label{3.9}
{T_H} = \frac{{\left( {d - 3} \right)\delta {p_i}}}{{4\pi }}
\end{equation}
The constant $\frac{{\left( {d - 3} \right)}}{{4\pi }}$ in the (3.9), is a calibration factor in d-dimensional space-time \cite{cavahli,cavahli1}. By a simple calculation based on equation (2.4), we obtain
\begin{equation}
\label{3.10}
\delta {p_i} = \left( {\frac{{\delta {x_i} + \alpha {l_p}}}{{4{\alpha ^2}l_p^2}}} \right)\left( {1 \pm \sqrt {\frac{{4{\alpha ^2}l_p^2\left( {1 + 4{\beta ^2}l_p^2{{\left( {\delta {x_i}} \right)}^2}} \right)}}{{{{\left( {\delta {x_i} + \alpha {l_p}} \right)}^2}}}} } \right)
\end{equation}
So, based on the above view point, the modified black hole Hawking temperature in a model universe with large extra dimension based on the ADD scenario and GUP becomes
\begin{equation}
\label{3.11}
T_d^{GUP} = \frac{{\left( {d - 3} \right)\left( {2{\zeta _d} + \alpha {l_p}} \right)}}{{16\pi {\alpha ^2}l_p^2}}\left[ {1 \pm \sqrt {1 - \frac{{4{\alpha ^2}l_p^2\left( {1 + 16{\beta ^2}l_p^2\zeta _d^2} \right)}}{{{{\left( {2{\zeta _d} + \alpha {l_p}} \right)}^2}}}} } \right]
\end{equation}
Therefore, the generalized uncertainty principle that admits a minimal length, a minimal momentum, and maximal momentum gives rise to the existence of a minimal mass, ${M_{\min }}$, of charged rotating TeV-scale black hole. In this way, if the negative sign is chosen in equation (3.11), the above result agrees with the standard result for large mass, based on the Heisenberg uncertainty principle. However, there is no evident physical meaning for the positive sign in equation (3.11).
\\ In order to find the concrete form of charged rotating micro black hole based on GUP, we consider the loss of information caused by a captured particle by the black hole which results in the increase of black hole entropy. We obtain
\begin{equation}
\label{3.12}
\delta S \simeq \frac{{dS}}{{dA}}\delta A.
\end{equation}
After some simple calculation, one can rewrite inequality (2.4) as a Heisenberg uncertainty principle format, $\delta {x_i}\delta {p_i} \ge \gamma \hbar '$ where ${\hbar '}$ may be regarded as an effective (modified) Planck constant \cite{xiang}. Correspondingly, the increase in area satisfies
\begin{equation}
\label{3.13}
\delta A \ge \gamma \hbar '
\end{equation}
where $\gamma  = 4\ln 2$ is a calibration factor.
\\ The black hole increase in entropy by $\delta {S_{\min }} = \ln 2$, as, the information of one bit is lost when a particle vanishes. On the other hand, the lower limit of (3.13) gives the minimum increase in the horizon area. So, we obtain
\begin{equation}
\label{3.14}
\begin{array}{l}
\frac{{dS}}{{dA}} \simeq \frac{{\delta {A_{\min }}}}{{\delta {S_{\min }}}} = \frac{{\gamma \hbar '}}{{\ln 2}}\\
= \frac{2}{{{\alpha ^2}l_p^2}}\left[ {\left( {2\zeta \alpha {l_p} + 4{\zeta ^2}} \right) - 2\zeta \sqrt {{{\left( {\alpha {l_p} + 2\zeta } \right)}^2} - 4{\alpha ^2}l_p^2\left( {1 + 16{\beta ^2}l_p^2{\zeta ^2}} \right)} } \right]
\end{array}
\end{equation}
Therefore, based on equation (3.14), the black hole entropy can be expressed as 
\begin{equation}
\label{3.15}
S \simeq \int {\frac{{\delta {S_{\min }}}}{{\delta {A_{\min }}}}dA = \frac{1}{4}\int {\frac{{dA}}{{\hbar '}}} } 
\end{equation}
From the Bekenstein-Smarr differential mass formula \cite{bekenstein1,smarr,smarr1}:
\begin{equation}
\label{3.16}
dM = \frac{1}{2}\kappa dA + \Phi dQ + \Omega dJ
\end{equation}
where $\kappa$, $\Phi$, and $\Omega$ denote the surface gravity, electrostatic potential of the event horizon, and angular velocity of Kerr-Newman black hole with conserved charge and angular momentum. It is easy to show that \cite{ghosh}
\begin{equation}
\label{3.17}
{A_H} = r_ + ^{d - 4}\zeta \frac{{2{\pi ^{\frac{{d - 1}}{2}}}}}{{\Gamma \left( {\frac{{d - 1}}{2}} \right)}}
\end{equation}
According to equations (3.13) and (3.14), $\zeta$ is a bridge which crosses the gap between A and S, hence it has geometric and thermodynamic meanings \cite{xiang}. Finally, we obtain the Bekenstein-Hawking entropy of charged rotating TeV-scale black hole in the presence of the generalized uncertainty principle cutoff effects as follows,
\begin{equation}
\label{3.18}
\begin{array}{l}
{S^{GUP}} = \frac{{{\pi ^{\frac{{d - 1}}{2}}}}}{{16\Gamma \left( {\frac{{d - 1}}{2}} \right)}} \times \\
\int {\frac{{\left[ {\left( {2\alpha {l_p}\sqrt {r_ + ^2 + {a^2}}  + 4r_ + ^2 + 4{a^2}} \right) + 2\left( {\sqrt {r_ + ^2 + {a^2}} } \right)\left( {\sqrt {{{\left( {\alpha {l_p} + 2\sqrt {r_ + ^2 + {a^2}} } \right)}^2} - 4{\alpha ^2}l_p^2\left( {1 + 16{\beta ^2}l_p^2\left( {r_ + ^2 + {a^2}} \right)} \right)} } \right)} \right]}}{{\left[ {1 + 16{\beta ^2}l_p^2\left( {r_ + ^2 + {a^2}} \right)} \right]}}}  \times \\
\left[ {\left( {d - 2} \right)r_ + ^{d - 3} + {a^2}\left( {d - 4} \right)r_ + ^{d - 5}} \right]d{r_ + }.
\end{array}
\end{equation}
It is easy to find that in the scenario of large extra dimensions, black hole entropy decreases and the classical picture breaks down since the black hole's entropy is small. So, the semi classical entropy could be used to measure the semi classical validity. 
\\We now proceed to discuss some thermodynamic and related black hole parameters in terms of different dimension cases.
\subsection{4-dimensions case}
When $d = 4$, we obtain $\Delta$ as follows (see 3.2)
\begin{equation}
\label{3.19}
\Delta  = {r^2} + {a^2} - 2Mr + {Q^2}
\end{equation}
which admits a solution ${r_ \pm }$, identified as the inner/outer event horizons. The outer event horizon is 
\begin{equation}
\label{3.19}
{r_ + } = M + \sqrt {\left( {{M^2} - {Q^2}} \right) - {a^2}}
\end{equation} 
In this case, the angular momentum J and charge Q must be restricted by ${\left( {\frac{J}{M}} \right)^2}\langle \left( {{M^2} - {Q^2}} \right)$. Other wise there is no horizon and one has a naked singularity. In which follow we calculate ${{T_d}^{GUP}}$ for $d = 4$ (outer four-dimensional brane)
\begin{equation}
\label{3.21}
\begin{array}{l}
T_4^{GUP} = \frac{1}{{16\left( {\left( {\alpha {l_p}M + 2\xi } \right)\pi {\alpha ^2}l_p^2} \right)}} \times \\
\left( {\alpha {l_p}M - {{\left( \begin{array}{l}
- 64{M^3}{\alpha ^2}{\beta ^2}l_p^4 - 64{J^2}{\alpha ^2}{\beta ^2}l_p^4 - 64M\sqrt {{M^4} - {M^2}{Q^2} - {J^2}} {\alpha ^2}{\beta ^2}l_p^4\\
- 3{M^2}{\alpha ^2}l_p^2 + 4M\xi \alpha {l_p} + 4{M^3} + 4{J^2} + 4\sqrt {{M^4} - {M^2}{Q^2} - {J^2}} 
\end{array} \right)}^{{\raise0.7ex\hbox{$1$} \!\mathord{\left/
{\vphantom {1 2}}\right.\kern-\nulldelimiterspace}
\!\lower0.7ex\hbox{$2$}}}} + 2\xi } \right)
\end{array}
\end{equation}
where $\xi  = \sqrt {{M^3} + {J^2} + 2\sqrt {{M^4} - {M^2}{Q^2} - {J^2}} }$. 
Based on the equation (3.11), the GUP gives rise to the existence of a minimal mass which one can find it by solving the following inequality
\begin{equation}
\label{3.22}
\frac{1}{4}\frac{{{{\left( {1 + 2\sqrt { - 12{\alpha ^2}{\beta ^2}l_p^4} } \right)}^2}l_p^2{\alpha ^2}}}{{{{\left( {16{\alpha ^2}{\beta ^2}l_p^4 - 1} \right)}^2}}}\langle M + \sqrt {{M^2} - {Q^2} - \frac{{{J^2}}}{{{M^2}}}}  + \frac{{{J^2}}}{{{M^2}}}
\end{equation} 
This inequality can be solved numerically. One can find easily that the Hawking temperature of a charged rotating micro black hole in 4-dimensions space-time is only defined for $M\rangle {M_{\max }}$.
\subsection{6-dimensions case}
Equation (3.6) for $d = 6$ reduces to
\begin{equation}
\label{3.23}
{r^3} + ({a^2} - {Q^2})r - \mu  = 0
\end{equation} 
which gives the event horizon radius as follows
\begin{equation}
\label{3.24}
r_ + ^6 = \frac{{{\chi ^{\frac{1}{3}}}}}{6} - \frac{{2\left( {{a^2} - {Q^2}} \right)}}{{{\chi ^{\frac{1}{3}}}}}
\end{equation}
where $\chi  = 108\mu  + 12\sqrt { - 12{Q^2} + 36{Q^4}{a^2} - 36{Q^2}{a^4} + 12{a^6} + 81{\mu ^2}}$ and according to the equation (3.3), $\mu  = 0.46M$. Obviously, the angular momentum and charge should be restricted to avoid no horizon and naked singularity. So, we obtain 
\begin{equation}
\label{3.25}
\begin{array}{l}
T_6^{GUP} = \frac{3}{{16\left( {\left( {3\alpha {l_p}M{\chi ^{\frac{1}{3}}} + \eta } \right)\pi {\alpha ^2}l_p^2} \right)}} \times \\
\left( {3\alpha {l_p}M{\chi ^{\frac{1}{3}}} - {{\left( \begin{array}{l}
- 2304{M^2}{Q^2}{\alpha ^2}{\beta ^2}l_p^4 + 4608{M^2}{Q^2}{a^2}{\alpha ^2}{\beta ^2}l_p^4 - 384{M^2}{\chi ^{\frac{2}{3}}}{Q^2}{\alpha ^2}{\beta ^2}l_p^4\\
- 2304{Q^2}{a^4}{\alpha ^2}{\beta ^2}l_p^4 + 384{M^2}{\chi ^{\frac{2}{3}}}{Q^2}{a^2}{\alpha ^2}{\beta ^2}l_p^4 - 16{M^2}{\chi ^{\frac{4}{3}}}{\alpha ^2}{\beta ^2}l_p^4 - \\
2304{\chi ^{\frac{2}{3}}}{J^2}{\alpha ^2}{\beta ^2}l_p^4 - 27{M^2}{\chi ^{\frac{2}{3}}}{\alpha ^2}l_p^2 + 6M\eta {\chi ^{\frac{1}{3}}}\alpha {l_p} + {M^2}{\chi ^{\frac{4}{3}}} + 24{M^2}{\chi ^{\frac{2}{3}}}{Q^2}\\
- 24{M^2}{\chi ^{\frac{2}{3}}}{a^2} + 144{M^2}{Q^4} - 288{M^2}{Q^2}{a^2} + 144{M^2}{a^4} + 144{J^2}{\chi ^{\frac{2}{3}}}
\end{array} \right)}^{\frac{1}{2}}} + \eta } \right)
\end{array}
\end{equation}
with $\eta  = {\left( {{M^2}{\chi ^{\frac{4}{3}}} + 24{M^2}{\chi ^{\frac{2}{3}}}{Q^2} - 24{M^2}{\chi ^{\frac{2}{3}}}{a^2} + 144{M^2}{Q^4} - 288{M^2}{a^2}{Q^2} + 144{M^2}{a^4} + 144{J^2}{\chi ^{\frac{2}{3}}}} \right)^{\frac{1}{2}}}$.
\\In the same way, one can also obtain the Hawking temperature and ${M_{\min }}$ respectively for other dimensions.
\\In the scenario with large extra dimensions based on the ADD model in the presence of the most general modified uncertainty principle, the temperature of the higher dimensional charged rotating black hole increases with respect to the dimensions in constant M (Figure 1).
\begin{figure}[tbp]
\centering % \begin{center}/\end{center} takes some additional vertical space
\includegraphics[width=1\textwidth,trim=-200 400 0 100,clip]{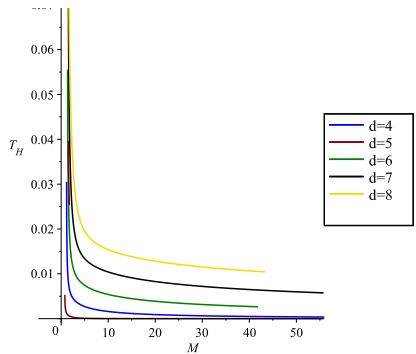}
\hfill
\caption{\label{fig1} Hawking temperature for different space-time dimensions.}
\end{figure}  
In fact, the higher dimensional black holes at fixed event horizon radii are hotter and their minimum mass as black hole remnant mass increase (see Figure 2).
\begin{figure}[tpb]
\centering \begin{center} \end{center} 
\includegraphics[width=1\textwidth,trim=-200 400 0 100,clip]{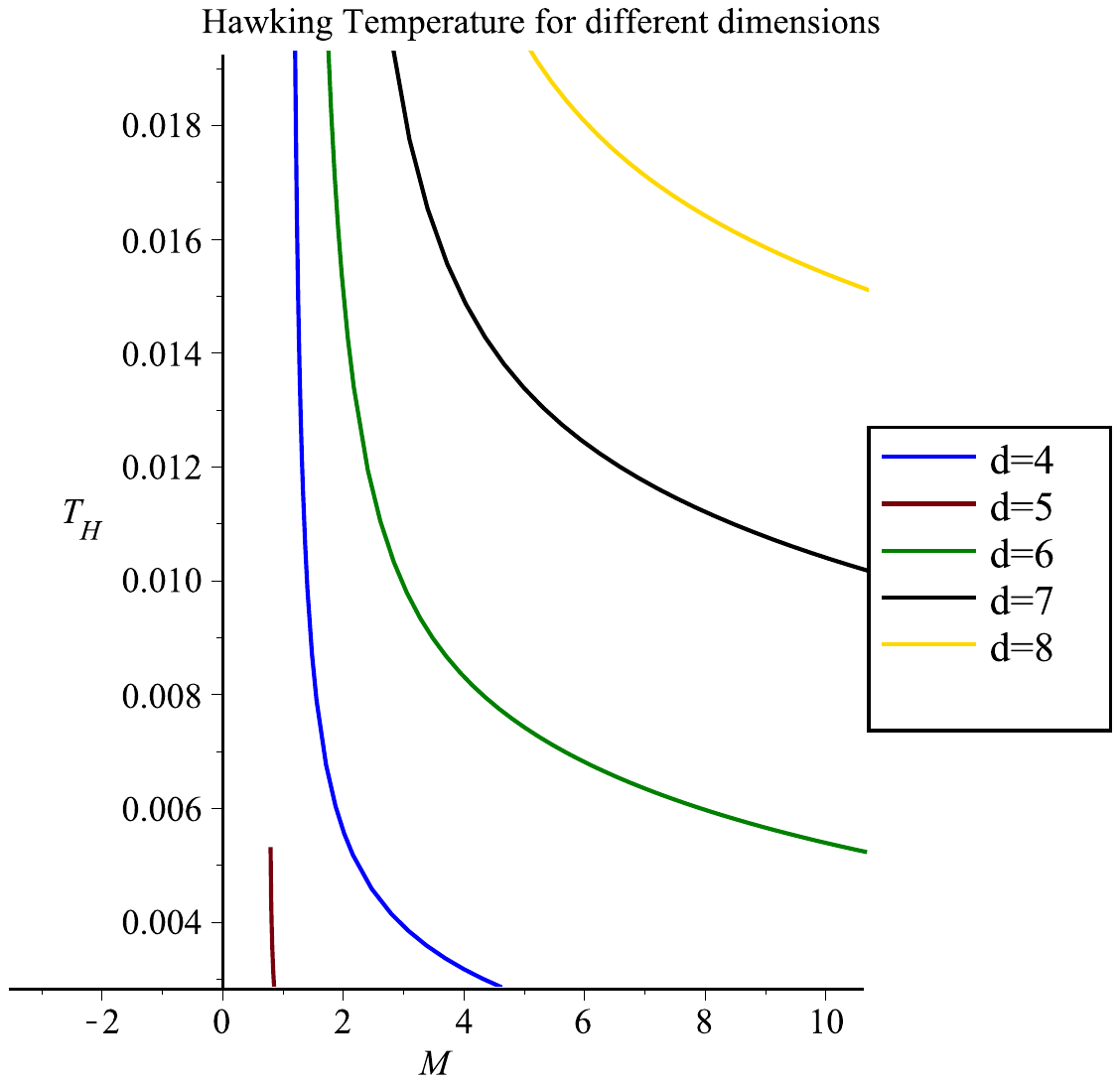}
\hfill
%\includegraphics[width=.45\textwidth,origin=c,angle=180]{img2.pdf}
% "\includegraphics" is very powerful; the graphicx package is already loaded
\caption{\label{fig2} Hawking temperature increases and shift vs. dimensions.}
\end{figure} 
In this case, there is restriction of the range of the parameters $\alpha$ and $\beta$, based on the equation (3.11) and its respective minimum mass which shows that $\alpha$ and $\beta$ cannot take arbitrary value. Meanwhile, $\alpha$ and $\beta$ are related parameters which  depend on the aspects of the candidates for quantum gravity proposal.
\section{Effects of angular momentum and charge on thermodynamics}
In order to characterize black hole, there are only mass, charge, and angular momentum \cite{padmanabhan,frolov}. Microscopic black holes are called thermal black hole. This type of black holes are expected to go through different stages during their life time \cite{giddings} as follows: I) The balding phase: at this initial stage, the black hole emits mainly gravitational radiations and sheds all the quantum numbers and multiple momenta apart from those determined by its mass, charge and angular momentum. II) The spin-down phase: during this stage, the black hole starts losing its angular momentum through the emission of Hawking radiation. III) The Schwarzschild phase: the black hole is no longer rotating and continues to lose its mass in the form of Hawking radiation. IV) The Planck phase: at this final stage the black hole mass approaches the true Planck scale as black hole remnant with mass ${M_{min}}$ which was discussed in the previous sections. The allowed initial particles forming TeV-scale black holes at the LHC are quarks, anti-quarks, and gluons which can form different types of the black holes in terms of charge and spin \cite{gingrich}. Although we are dealing with a micro black hole as a semi-classical black hole, for simplicity we have considered the initial angular momentum of the black hole due to the spin states of the incoming partons and ignore the possibility of an initial small orbital angular momentum due to an impact parameter.  
\\ Figure 2 shows the angular momentum effects on the minimum mass and maximum Hawking temperature. The minimum mass and its order of magnitude increases when angular momentum increases and meantime the temperature peak displaces to the lower temperature (see Figure 3).
\begin{figure}[tpb]
\centering \begin{center} \end{center} 
\includegraphics[width=1\textwidth,trim=-200 400 0 100,clip]{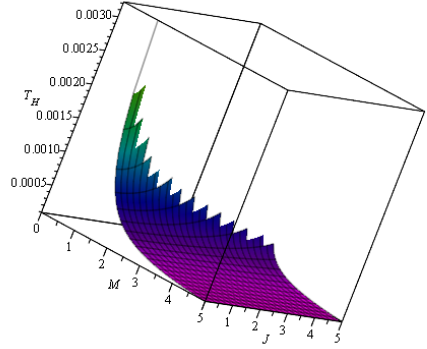}
\hfill
%\includegraphics[width=.45\textwidth,origin=c,angle=180]{img2.pdf}
% "\includegraphics" is very powerful; the graphicx package is already loaded
\caption{\label{fig3} Spin-down phase effect on the black hole temperature.}
\end{figure}   
After the balding and spin-down phases, the black hole will decay via the semi-classical Hawking evaporation process \cite{hawking}. We have shown that after spin-down phase the black hole continues to evaporate without rotation which is considered as the Schwarzschild phase (Figures 4 and 5).
\\For the spin-down phase, the absorption probability also depends on the angular momentum parameter and the space-time properties and emission rate increases with the increase in the angular momentum \cite{ido}. 
Figures 3 and 4 show no special distribution in terms of the angular momentum during the spin-down phase because this phase has a preferred axis for the bran localized emission, the rotation axis of the TeV-scale black hole \cite{kanti}.
\begin{figure}[tpb]
\centering \begin{center} \end{center} \includegraphics[width=1\textwidth,trim=-200 400 0 100,clip]{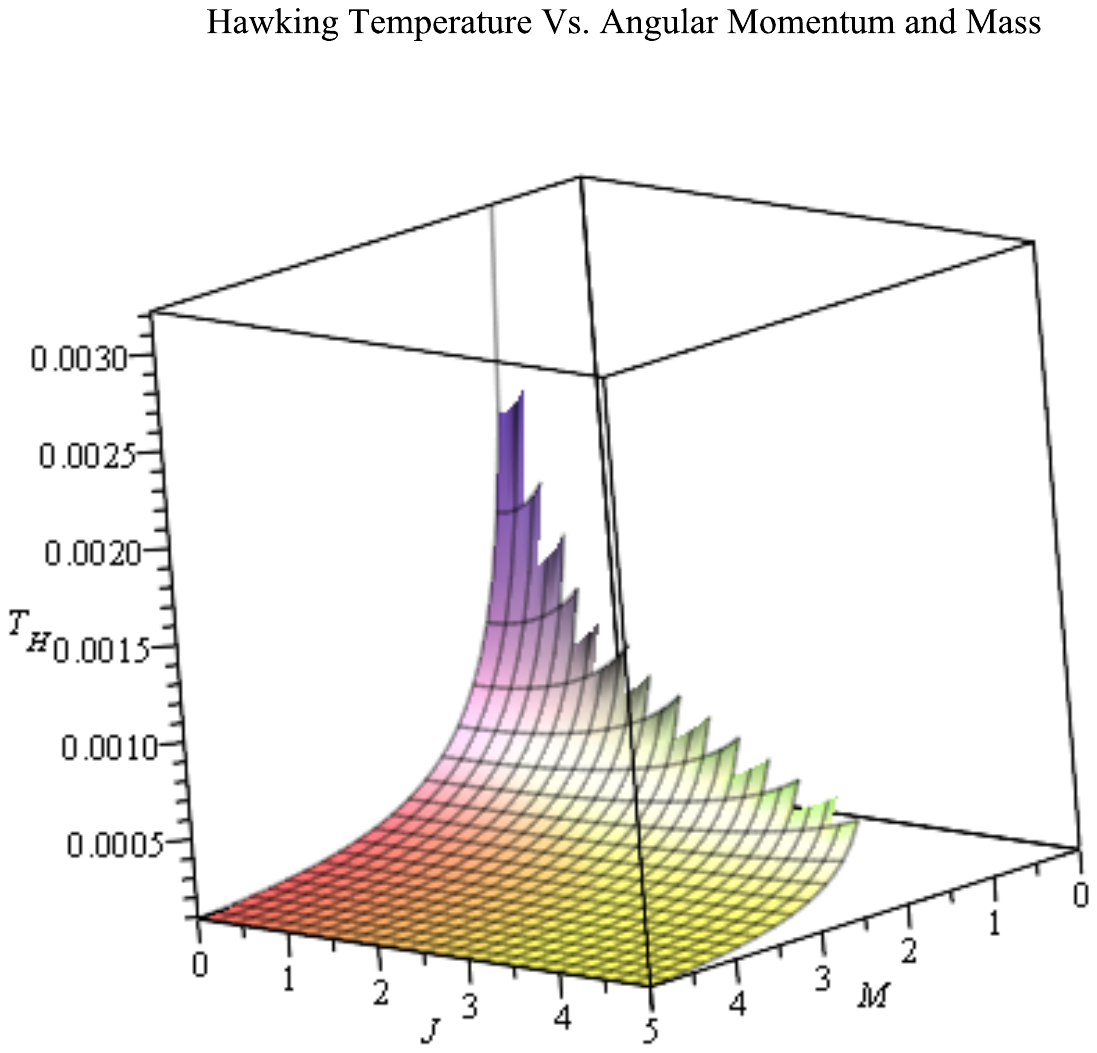}
\hfill
%\includegraphics[width=.45\textwidth,origin=c,angle=180]{img2.pdf}
% "\includegraphics" is very powerful; the graphicx package is already loaded
\caption{\label{fig4} Angular momentum, J, and mass, M vs. the Hawking temperature.}
\end{figure}
\begin{figure}[tpb]
\centering \begin{center} \end{center} 
\includegraphics[width=1\textwidth,trim=-200 400 0 100,clip]{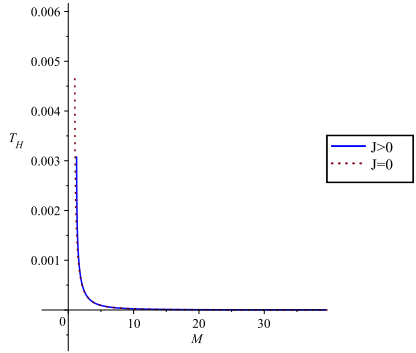}
\hfill
%\includegraphics[width=.45\textwidth,origin=c,angle=180]{img2.pdf}
% "\includegraphics" is very powerful; the graphicx package is already loaded
\caption{\label{fig5} Continue evaporation after the spin-down phase.}
\end{figure}
The initial particles formed nine possible charge states: $\pm \frac{4}{3}, \pm 1, \pm \frac{2}{3}, \pm \frac{1}{3},0$ \cite{landsberg}. Figure 6 shows the charge effects on the minimum mass and maximum Hawking temperature of black hole. Obviously, when charge increases, the minimum mass and its order of magnitude increases, and the temperature peak displaces to the lower temperature as well as angular momentum effect. Here, we would like to draw attention that we have described the higher dimensional Kerr-Newman black hole geometry based on the solution of the Einstein equation field in Ref. \cite{ghosh} due to the line element dependency to gauge charge compared to the proposed solution in \cite{aliev}. Although it was shown \cite{dai} that the TeV-scale black hole charge will reach zero much faster than its mass and the charge to mass ratio is much less than one \cite{abbasvandi}, we have shown here that charge is more effective on Hawking temperature variation compared to the angular momentum in fixed mass (see Figure 7).
\\Therefore, if the fundamental Planck scale is of the order of TeV, the Large Hadron Collider would produce charged rotating micro black hole which as a consequence of their evaporation, degenerate to charged non-rotating TeV-scale balck hole at the last stage and thus yield charged black hole remnants.
\section{Frame dragging and Sagnac effect}
The rotation in space-time has became the source of stimulating and fascinating physical issues for a couple of decades. The interferometrical detection of the effects of rotation has a long story dating back to the end of the XIX century which can be found in \cite{anderson}. In 1913, Sagnac predicted \cite{sagnac,sagnac1} a fringe shift for monochromatic light waves in vacuum, counter propagating along a closed path in a rotating interferometer. In fact, Sagnac in the first reported an experimental observation of the effect of rotation in space-time. The Sagnac effect has been detected in many experiments \cite{anandan,razzi,rodrigues}, and it can be generalized in the framework of general relativity. The standard Sagnac effect dynamics, in the background of natural splitting can be described in terms of a gravito-electromagnetic formal analogy \cite{cattaneo}, which leads to the formulation of the gravito-magnetic Aharonov-Bohm effect. It has been shown \cite{rizzi,rizzi1,ruggiero} that the Sagnac phase shift and the time delay, as effects of rotation on space-time, can be obtained by the above mentioned analogy.
\\We investigate the Sagnac effect in the higher dimentional Kerr-Newman fields in the presence of GUP considered in previous sections as a potential solution of a microscopic charged rotating black hole. By generalizing the conclusion of Rizzi and Ruggiero \cite{rizzi,rizzi1,ruggiero} for both the matter and light beams, counter-propagating on a round trip in a interferometer rotating in both flat and curved space-time, the GM Aharonov-Bohm phase shift is as follows,
\begin{equation}
\label{5.1}
\Delta \Phi  = \frac{{2m{\varsigma _0}}}{{c\hbar }}\oint\limits_c {{{\tilde A}_G}d\tilde x = \frac{{2m{\varsigma _0}}}{{c\hbar }}\oint\limits_s {\tilde B{}_Gd\tilde s} }
\end{equation}
\begin{figure}[tpb]
\centering \begin{center} \end{center} 
\includegraphics[width=1\textwidth,trim=-200 400 0 100,clip]{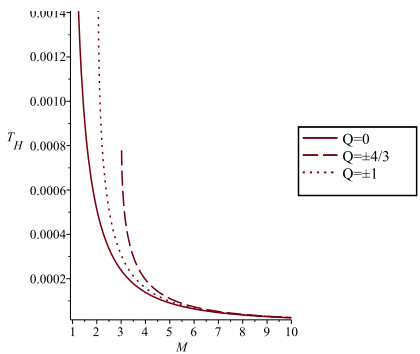}
\hfill	%\includegraphics[width=.45\textwidth,origin=c,angle=180]{img2.pdf}
% "\includegraphics" is very powerful; the graphicx package is already loaded
\caption{\label{fig6} Effect of different charge amounts on Hawking temperature}
\end{figure}    
Here, the parameter $m$ is the standard relative mass of particle moving in a uniformly rotating frame. The proper time difference corresponding to (5.1) is obtained as 
\begin{equation}
\label{5.2}
\Delta T = \frac{{2{\varsigma _0}}}{{{c^3}}}\oint\limits_c {{{\tilde A}_G}} d\tilde x = \frac{{2{\varsigma _0}}}{{{c^3}}}\oint\limits_s {{{\tilde B}_G}d\tilde s}
\end{equation}
which can be measured by an observer provided with a standard clock. Equations (5.1) and (5.2) are based on the hypothesis that the two beams have locally the same relative velocity in the reference frame defined by the congruence $\Gamma$, and the particles are supposed to neglect their spins. The metric we deal with is given in coordinate adapted to a congruence of asymptotically inertial observers. So, we should define the time like congruence of rotating observers. In this case, let us introduce the coordinate transformation
\begin{equation}
\label{5.3}
\begin{array}{l}
ct = ct'\\
r = r'\\
\varphi  = \varphi ' - {\Omega _d}t'\\
\theta  = \theta '
\end{array}
\end{equation}
where $c$ is constant and $\Omega _d$ is angular velocity. The above transformation defines the passage from a chart adapted to the inertial frame to a chart adapted to the rotating frame \cite{rizzi1}.  
Here, ${\Omega _d}$, is the angular velocity of the observer . Now, one can obtain the space-time metric in coordinates adapted to congruence of rotating observers and calculate the unit vectors field $\varsigma \left( x \right)$,
\begin{equation}
\label{5.4}
{\varsigma _i} = {g_{i0}}{\varsigma ^0}
\end{equation}
and the graviro-magnetic vector potential
\begin{equation}
\label{5.5}
\tilde A_i^G \approx {c^2}\frac{{{\varsigma _i}}}{{{\varsigma _0}}}
\end{equation}
\begin{figure}[tpb]
\centering \begin{center} \end{center} 	\includegraphics[width=1\textwidth,trim=-200 400 0 100,clip]{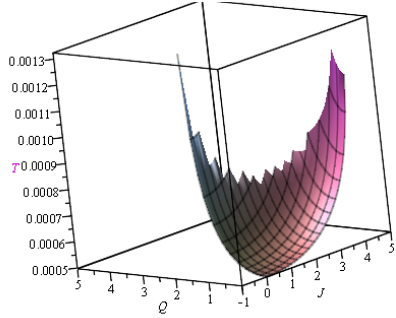}
\hfill	%\includegraphics[width=.45\textwidth,origin=c,angle=180]{img2.pdf}
% "\includegraphics" is very powerful; the graphicx package is already loaded
\caption{\label{fig7} Effect of the angular momentum, J and charge, Q on the Hawking temperature}
\end{figure}
In order to investigate the Sagnac effect as a frame dragging effect at the quantum gravity regime for charged rotating TeV-scale black holes, we consider higher dimensional Kerr-Newman space-time in the presence of GUP. We apply the above described formalism on relation (3.1) and consider the counter-propagating beams around a charged rotating micro black hole, in its equatorial plane along circular space trajectories by setting $r = R = const$. We use again the units such that $G = c = \hbar = 1$. Applying the transformation (5.3) to the metric (3.1) and setting $\theta  = \frac{\pi }{2}$, we obtain
\begin{equation}
\label{5.6}
\begin{array}{l}
d{s^2} = \left[ {1 - \frac{\mu }{{{r^{d - 3}}}} - \frac{{{Q^2}}}{{{r^2}}} + 2a{\Omega _d}\left( {\frac{\mu }{{{r^{d - 3}}}} - \frac{{{Q^2}}}{{{r^2}}}} \right) - \Omega _d^2\left[ {{r^2} + {a^2}\left( {1 + \frac{\mu }{{{r^{d - 3}}}} + \frac{{{Q^2}}}{{{r^2}}}} \right)} \right]} \right]dt\\
- \frac{{{r^2}d{r^2}}}{{{r^2} + {a^2} - \frac{\mu }{{{r^{d - 5}}}} - \frac{{{Q^2}}}{{{r^2}}}}} + \left[ {2a\left( {\frac{\mu }{{{r^{d - 3}}}} - \frac{{{Q^2}}}{{{r^2}}}} \right) - 2{\Omega _d}\left[ {{r^2} + {a^2}\left( {1 + \frac{\mu }{{{r^{d - 3}}}} - \frac{{{Q^2}}}{{{r^2}}}} \right)} \right]} \right]dtd\varphi \\
- \left[ {{r^2} + {a^2}\left( {1 + \frac{\mu }{{{r^{d - 3}}}} + \frac{{{Q^2}}}{{{r^2}}}} \right)} \right]d{\varphi ^2}
\end{array}
\end{equation}
By obtaining the components of the vector field $\varsigma \left( x \right)$, for the gravito-magnetic potential, we have
\begin{equation}
\label{5.7}
\begin{array}{l}
\tilde A_3^G \approx \left[ {2{\Omega _d}\left[ {{R^2} + {a^2}\left( {1 + \frac{\mu }{{{R^{d - 3}}}} - \frac{{{Q^2}}}{{{R^2}}}} \right)} \right] - 2a\left( {\frac{\mu }{{{R^{d - 3}}}} - \frac{{{Q^2}}}{{{R^2}}}} \right)} \right] \times \\
{\left[ {\Omega _d^2\left[ {{R^2} + {a^2}\left( {1 + \frac{\mu }{{{R^{d - 3}}}} + \frac{{{Q^2}}}{{{R^2}}}} \right)} \right] - 2a{\Omega _d}\left( {\frac{\mu }{{{R^{d - 3}}}} - \frac{{{Q^2}}}{{{R^2}}}} \right) + \frac{\mu }{{{R^{d - 3}}}} + \frac{{{Q^2}}}{{{R^2}}} - 1} \right]^{ - 1}}
\end{array}
\end{equation}
Substituting (5.7) into equation (5.1) gives
\begin{equation}
\label{5.8}
\begin{array}{l}
\Delta \Phi  = 4\pi m\left[ {2{\Omega _d}\left[ {{R^2} + {a^2}\left( {1 + \frac{\mu }{{{R^{d - 3}}}} - \frac{{{Q^2}}}{{{R^2}}}} \right)} \right] - 2a\left( {\frac{\mu }{{{R^{d - 3}}}} - \frac{{{Q^2}}}{{{R^2}}}} \right)} \right] \times \\
{\left[ {\Omega _d^2\left[ {{R^2} + {a^2}\left( {1 + \frac{\mu }{{{R^{d - 3}}}} + \frac{{{Q^2}}}{{{R^2}}}} \right)} \right] - 2a{\Omega _d}\left( {\frac{\mu }{{{R^{d - 3}}}} - \frac{{{Q^2}}}{{{R^2}}}} \right) + \frac{\mu }{{{R^{d - 3}}}} + \frac{{{Q^2}}}{{{R^2}}} - 1} \right]^{{\raise0.7ex\hbox{${ - 1}$} \!\mathord{\left/
{\vphantom {{ - 1} 2}}\right.\kern-\nulldelimiterspace}
\!\lower0.7ex\hbox{$2$}}}}
\end{array}
\end{equation}
On the other hand, for a photon, we can introduce the standard relative mass 
\begin{equation}
\label{5.9}
m = \varepsilon = h\nu
\end{equation}
\begin{figure}[tpb]
\centering \begin{center} \end{center} 
\includegraphics[width=1\textwidth,trim=-200 400 0 100,clip]{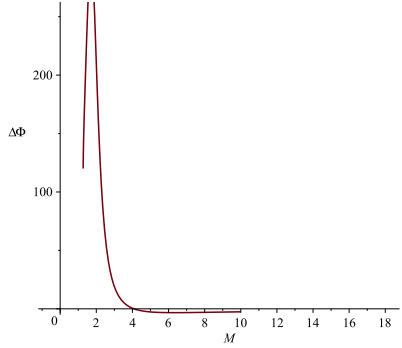}
\hfill	%\includegraphics[width=.45\textwidth,origin=c,angle=180]{img2.pdf}
% "\includegraphics" is very powerful; the graphicx package is already loaded
\caption{\label{fig8} The phase shift Vs. mass variation}
\end{figure}      
Incorporating GUP, the de Broglie relation is given by 
\begin{equation}
\label{5.10}
\lambda  = \frac{1}{p}\left( {1 - \alpha {l_p}p + {\alpha ^2}l_p^2{p^2}} \right)\left( {1 + \frac{{{\beta ^2}l_p^2}}{{{p^2}}}\left( {1 - \alpha {l_p}p + {\alpha ^2}l_p^2{p^2}} \right)} \right)
\end{equation}
or equivalently
\begin{equation}
\label{5.11}
\varepsilon  = E\left( {1 - \alpha {l_p}E + {\alpha ^2}l_p^2{E^2}} \right)\left( {1 + \frac{{{\beta ^2}l_p^2}}{{{E^2}}}\left( {1 - \alpha {l_p}E + {\alpha ^2}l_p^2{E^2}} \right)} \right)
\end{equation}
As a consequence by substituting (5.11) into (5.8), one can obtain the phase shift of equation (5.1) as follows
\begin{equation}
\label{5.12}
\begin{array}{l}
\Delta \Phi  = 4\pi E\left( {1 - \alpha {l_p}E + {\alpha ^2}l_p^2{E^2}} \right)\left( {1 + \frac{{{\beta ^2}l_p^2}}{{{E^2}}}\left( {1 - \alpha {l_p}E + {\alpha ^2}l_p^2{E^2}} \right)} \right) \times \\
\left[ {2{\Omega _d}\left[ {{R^2} + {a^2}\left( {1 + \frac{\mu }{{{R^{d - 3}}}} - \frac{{{Q^2}}}{{{R^2}}}} \right)} \right] - 2a\left( {\frac{\mu }{{{R^{d - 3}}}} - \frac{{{Q^2}}}{{{R^2}}}} \right)} \right] \times \\
{\left[ {\Omega _d^2\left[ {{R^2} + {a^2}\left( {1 + \frac{\mu }{{{R^{d - 3}}}} + \frac{{{Q^2}}}{{{R^2}}}} \right)} \right] - 2a{\Omega _d}\left( {\frac{\mu }{{{R^{d - 3}}}} - \frac{{{Q^2}}}{{{R^2}}}} \right) + \frac{\mu }{{{R^{d - 3}}}} + \frac{{{Q^2}}}{{{R^2}}} - 1} \right]^{{\raise0.7ex\hbox{${ - 1}$} \!\mathord{\left/
{\vphantom {{ - 1} 2}}\right.\kern-\nulldelimiterspace}
\!\lower0.7ex\hbox{$2$}}}}
\end{array}
\end{equation}
and accordingly for the time delay of equation (5.12), we obtain
\begin{equation}
\label{5.13}
\begin{array}{l}
\Delta \Phi  = 4\pi \left[ {2{\Omega _d}\left[ {{R^2} + {a^2}\left( {1 + \frac{\mu }{{{R^{d - 3}}}} - \frac{{{Q^2}}}{{{R^2}}}} \right)} \right] - 2a\left( {\frac{\mu }{{{R^{d - 3}}}} - \frac{{{Q^2}}}{{{R^2}}}} \right)} \right] \times \\
{\left[ {\Omega _d^2\left[ {{R^2} + {a^2}\left( {1 + \frac{\mu }{{{R^{d - 3}}}} + \frac{{{Q^2}}}{{{R^2}}}} \right)} \right] - 2a{\Omega _d}\left( {\frac{\mu }{{{R^{d - 3}}}} - \frac{{{Q^2}}}{{{R^2}}}} \right) + \frac{\mu }{{{R^{d - 3}}}} + \frac{{{Q^2}}}{{{R^2}}} - 1} \right]^{{\raise0.7ex\hbox{${ - 1}$} \!\mathord{\left/
{\vphantom {{ - 1} 2}}\right.\kern-\nulldelimiterspace}
\!\lower0.7ex\hbox{$2$}}}}
\end{array}
\end{equation} 
According to equations (5.12) and (5.13), we find that the Sagnac effect depends on angular velocity, the path of the beams of the observers, the angular momentum, charge, the gravitational mass, and extra dimensions. Also, the phase shift depends on space-time structure in presence of the GUP. One can find that if $\Omega_d$ vanishes, a phase shift appears due to the source of gravitational field properties such as its angular momentum, mass, etc. Obviously, a charged rotating micro black hole could produce micro gravitational waves due to the mass variation in different phases of the Hawking radiation which needs more investigation (Figure 8).
\section{Conclusion}
In this paper, we have calculated the Hawking temperature (equation 3.11) and the Bekenstein-Hawking entropy (equation 3.18) of a charged rotating TeV-scale black hole in the framework of the extra dimensions scenario based on the ADD model through a most general form of the GUP as our primary input. In such an extra dimensional scenario in the presence of the GUP, the black hole temperature increases with respect to the dimensions (Figure 1). It is evident that in large extra dimensions the higher dimensional black hole at fixed event horizon radii is hotter and its minimum mass increases (Figure 2) which is more than its four dimensional counterpart. It yields charged rotating black hole remnant. Also, the angular momentum affects the minimum mass of the black hole and maximum temperature. We have shown that the minimum mass and its order of magnitude increases when angular momentum increases while the temperature peak shift to the lower temperature (Figures 3 and 4). 
\\The black hole evaporation continues after the spin-down phase without any rotation which is considered as the Schwarzschild phase (see Figure 5). Additionally, the electric charge affects on minimum mass of the black hole and maximum Hawking temperature as well. It is also shown that the minimum mass and its order of magnitude increase and the temperature peak shift to lower temperature when the electric charge increases (Figure 6). It is in full agreement with the results of previous manuscript of authors \cite{abbasvandi}. We have shown that the charge variation is more effective on the temperature compared to the angular momentum variation (see Figure 7). We have also calculated the phase shift and time delay related to the frame dragging and Sagnac effect of charged rotating TeV-Scale black hole.

\section*{Aknowledgement}
The paper is supported by University of Malaya (Grant No. Ru-023-2014) and Universiti Kebangsaan Malaysia (Grant No. FRGS/2/2013/ST02/UKM/02/2).


\end{document}